\documentclass[twocolumn,preprintnumbers,amsmath,amssymb,superscriptaddress]{revtex4-2}
\UseRawInputEncoding

\usepackage{latexsym,amssymb,amsthm,amsmath,epsfig}
\usepackage[left=2.00cm, right=2.00cm, top=2.00cm, bottom=2.00cm]{geometry}
\usepackage{float}
\usepackage{hyperref}
\hypersetup{
    citecolor=red,
    colorlinks=true,
    linkcolor=blue,
    filecolor=blue,      
    urlcolor=blue,
}

\begin{document}

\title{Elastic collision rates of spin-polarized fermions in two dimensions}

%%%%%%%%%%%%%%%%%%%%%%%%%%%%%%%% Authors names and their affiliations %%%%%%%%%%%%%%%%%%%%%%%%%%%%%%%%%%%%%%%%%%%%%%%%%%%%

\author{Muhammad Awais Altaf}
%\email{mwaseem@pieas.edu.pk}
\affiliation{Department of Physics and Applied Mathematics, Pakistan Institute of Engineering and Applied Science (PIEAS), Nilore, Islamabad 45650, Pakistan}
\affiliation{Department of Physics, University of Mianwali, Mianwali 42200, Pakistan}

\author{Takashi Mukaiyama}
\affiliation{Department of Physics, Tokyo Institute of Technology, 2-12-1 O-okayama, Meguro, Tokyo 152-8550, Japan}

\author{Muhammad Waseem}
\email{mwaseem@pieas.edu.pk}
\affiliation{Department of Physics and Applied Mathematics, Pakistan Institute of Engineering and Applied Science (PIEAS), Nilore, Islamabad 45650, Pakistan}
%\affiliation{Department of Physics and Applied Mathematics, PIEAS, Nilore, Islamabad 45650, Pakistan}
\affiliation{Center for Mathematical Sciences, PIEAS, Nilore, Islamabad 45650, Pakistan}
%%%%%%%%%%%%%%%%%%%%%%%%%%%%%%%%%%%%%%%%%%%%%%%% END %%%%%%%%%%%%%%%%%%%%%%%%%%%%%%%%%%%%%%%%%%%%%%%%%%%%%%%%%%%%%%%%%%%%%%%%%

%%%%%%%%%%%%%%%%%%%%%%%%%%%%%%%%%%%%%%%%%%%%%%%%%%%%%%%%%%%%%%%%%%
%%%%%%%%%%%%%%%%%%%%%%%%%%%%%%%%%%%%%%%%%
%\author{Muhammad Awais Altaf$^{1,2}$}
%\thanks{Present Address: Department of Physics, University of Mianwali, Mianwali 42200, Pakistan,}
%\email{mwaseem@pieas.edu.pk}
%\author{J $^{1,3}$}
%\author{T $^{1,3}$}
%\author{Muhammad Waseem$^{1,3}$}
%\affiliation{%
%$^{1}$\mbox{Department of Physics and Applied Mathematics, PIEAS, Nilore, Islamabad 45650, Pakistan,}\\
%$^{2}$\mbox{Department of Physics, University of Mianwali, Mianwali 42200, Pakistan,}\\
%$^{2}$\mbox{Center for Mathematical Sciences, PIEAS, Nilore, Islamabad 45650, Pakistan}\\
%$^{3}$\mbox{Graduate School of Engineering Science, Osaka University, 1-3, Machikaneyama, Toyonaka, Osaka 560-8531 Japan}\\
%}
%%%%%%%%%%%%%%%%%%%%%%%%%%%%%%%%%%%%%%%
%%%%%%%%%%%%%%%%%%%%%%%%%%%%%%%%%%%%%%%%%%%%%%%%%%%%%%%%%%%%%%%%%%%%

\date{\today }

\begin{abstract}
We study the $p$-wave elastic collision rates in a two-dimensional spin-polarized ultracold Fermi gas in the presence of a $p$-wave Feshbach resonance.
We derive the analytical relation of the elastic collision rate coefficient in the close vicinity of resonance when the effective range is dominant.
The elastic collision rate is enhanced by an exponential scaling of $e^{-q_{r}^{2} / q_{T}^{2}}$ towards the resonance.
 Here, $q_{r}$ is the resonant momentum and $q_T$ is the thermal  momentum. 
An analogous expression derived for the case of three dimensions successfully explains the thermalization rates measurement in the recent experiment~[Phys. Rev. A 88, 012710 (2013)].  
In the zero-range limit where the effective range is negligible, the elastic collision rate coefficient is proportional to temperature $T^2$ and scattering area $A_{p}^2$. 
Using these elastic collision rates we studied the energy transfer from high to low velocity through $p$-wave collision. We also discuss the collisional stability in the presence of three-body losses in the background scattering limit.
Our results may provide insight into the dynamics of the two-dimensional spin-polarized ultracold Fermi gas.
\end{abstract}

\maketitle

%==========================================================================================================
\section{introduction}
%==========================================================================================================
Placing spin-polarized fermions into the lowest identical hyperfine ground state forbids the $s$-wave scattering and leaves $p$-wave scattering as a dominant scattering channel. 
The $p$-wave interactions are very weak at low temperatures in single-component ultracold Fermi gases~\cite{demarco_measurement_1999}.
However, interactions between two identical fermions can be enhanced using the $p$-wave Feshbach resonances and have been observed in some experiments~\cite{zhang_p-wave_2004,regal_tuning_2003,schunck_feshbach_2005,gerken_observation_2019,ticknor_multiplet_2004}.
In the past, $p$-wave Feshbach resonances have been utilized for Feshbach molecules creation~\cite{waseem_creation_2016, gaebler_p-wave_2007,fuchs_binding_2008,maier_radio-frequency_2010}, dimer association spectroscopy~\cite{ahmed-braun_probing_2021},
non-resonant light control of the $p$-wave Feshbach resonances~\cite{crubellier_controlling_2019}, Fermi liquid properties~\cite{ding_fermi-liquid_2019}, normal-state properties~\cite{yao_normal-state_2018}, few-body physics~\cite{schmidt_three-body_2020, waseem_unitarity-limited_2018,yoshida_scaling_2018,suno_recombination_2003, higgins_three_2022, zhu_three-body_2022,esry_threshold_2001}, out-of-equilibrium thermodynamics~\cite{luciuk_evidence_2016,yu_universal_2015,yoshida_universal_2015}, ground state energy and properties~\cite{liu_studies_2018, bertaina_quantum_2023} and elastic unitary $p$-wave interactions~\cite{venu_unitary_2023}.

Single-component Fermi gases are of fundamental interest such as Kohn-Luttinger instabilities at weak interactions~\cite{kohn_new_1965}, unconventional superfluidity~\cite{kojima_spin_2008, leduc_spin_1990,fedorov_p-wave_2017}, shifts of atomic clock frequency~\cite{lemke_p-wave_2011}, rich quantum phase diagrams~\cite{gurarie_quantum_2005}.
Such fundamental interests lead to great efforts in theoretical and experimental studies to understand elastic~\cite{demarco_measurement_1999, nakasuji_experimental_2013, top_spin-polarized_2021} and inelastic scattering processes close to $p$-wave Feshbach resonances~\cite{schmidt_three-body_2020, waseem_unitarity-limited_2018,yoshida_scaling_2018,suno_recombination_2003, higgins_three_2022, zhu_three-body_2022,esry_threshold_2001,welz_anomalous_2023}.
Elastic collisions are required to reach quantum degeneracy via evaporation.
For single-component Fermi gases, elastic collisions and thermalization have been measured near $p$-wave Feshbach resonances~\cite{nakasuji_experimental_2013, demarco_measurement_1999}.
Recently, thermalization and evaporative cooling by background $p$-wave collision have been observed for the three-dimensional case with modest efficiency~\cite{top_spin-polarized_2021}.

The $p$-wave Feshbach resonances not only enhance the binary elastic collision but also the inelastic collisions~\cite{schmidt_three-body_2020}.
Earlier experimental studies showed that spin-polarized fermions close to $p$-wave resonances are unstable, with a much short lifetime, due to inelastic collisions\cite{gaebler_p-wave_2007, inada_collisional_2008}. 
This shorter lifetime has led to theoretical studies to explore the reduction of inelastic collision losses in reduced dimensions~\cite{levinsen_stability_2008, kurlov_two-body_2017, fedorov_p-wave_2017}. 
On the fundamental side, fermions confined in two dimensions exhibit topological superfluid phases~\cite{read_paired_2000,ivanov_non-abelian_2001,yang_tricritical_2020} and few-body bound states~\cite{nishida_super_2013}.
However, understanding the elastic collision rates of spin-polarized fermions in two dimensions is also important to explore the possibility of evaporation~\cite{top_spin-polarized_2021}, non-equilibrium dynamics of system~\cite{luciuk_evidence_2016,hu_resonantly_2019}, hydrodynamics~\cite{gehm_unitarity-limited_2003} and quantum transport properties~\cite{hausler_interaction-assisted_2021,maki_transport_2023}.

In this paper, we focus on the $p$-wave elastic collision rates in a two-dimensional spin-polarized ultracold Fermi gas in the presence of a $p$-wave Feshbach resonance.
We consider two extreme regimes of $p$-wave resonance. 
One is the non-zero-range limit in the close vicinity of resonance where the effective range term is dominant. 
We derive the analytical expression of elastic collision rates in this regime, which is in agreement with direct numerical simulations. 
Our analytical expression shows that elastic collision rates are enhanced exponentially towards the resonance.
We have also derived an analogous analytical expression for a three-dimensional case that explains the experimental results of thermalization rates measured by Nakasuji \textit{et al}., ~\cite{nakasuji_experimental_2013}. 
The other regime is the zero-range limit where the effective range is negligible and the scattering area $A_p$ is near the background $p$-wave collision in the ultracold regime.
In this zero-range limit, the elastic collision rate coefficient is proportional to temperature $T^2$ and scattering area $A_{p}^2$. 
We also analyze the transfer of energy from high to low velocities through $p$-wave collision near the background $p$-wave collision and as well as close to the resonance. 
%Our analysis shows that the transfer of energy is $\sqrt{2}$ times faster than the three-dimensional case. This suggests that $p$-wave evaporative cooling in two dimensions can be performed better than recently achieved modest efficiency in three dimensions~\cite {top_spin-polarized_2021}.
We also discuss the collisional stability in the presence of three-body losses.
We show that in the region of background $p$-wave collision, the ratio of good-to-bad collision rates can be improved as compared to the three-dimensional case. 

This article is organized as follows. Section II describes our theoretical analysis of elastic collision rates. 
In Sec. III, we describe energy transfer analysis.   
In Sec. IV, we explore the elastic-to-inelastic collision ratio, and finally, we concluded.

%===========================================================================================================

%%%%%%%%%%%%%%%%%%%%%%%%%%%%%%%%%%%%%%%%%%%%%%%%%%%%%%%%%%%%%%%%%%%%%%

%==========================================================================================================
\section{Elastic Collision Rates}
%==========================================================================================================
Let's first consider two identical fermions colliding in quasi-two dimensions in the presence of $p$-wave Feshbach resonance.
the Quasi-two dimensions geometry can be obtained by tight harmonic confinement in the axial direction ($z$) with frequency $\omega_0$~\cite{waseem_creation_2016,gunter_p-wave_2005}. In the axial direction, the extension of the wave function is given by harmonic oscillator length $l_{0}=\sqrt{\hbar/m\omega_0}$. The interatomic separation $\rho$ in the plane $x,y$ greatly exceeds the harmonic oscillator length $l_0$.
Then $p$-wave relative motion is described by the wavefunction \cite{pricoupenko_resonant_2008, kurlov_two-body_2017},
\begin{equation}\label{psi}
    \psi_{2D}(\rho)=\varphi_{2D}(\rho)e^{\dot{\iota}\theta}\frac{1}{\left(2\pi l_0^2\right)^{1/4}} \exp \left(\frac{-z^2}{4l_0^2}\right);\hspace{0.25cm}\rho \gg l_0,
\end{equation}
where $\theta$ is the scattering angle, and $z$ is the inter-particle separation in the axial direction.
Considering the ultracold limit with respect to the axial motion we assume that confinement length $l_0$ is much larger than the characteristic radius of interaction $R_{e}$~\cite{pricoupenko_resonant_2008}.
Then two-dimension radial wave function for $p$-wave becomes \cite{pricoupenko_resonant_2008,kurlov_two-body_2017}
\begin{equation}\label{psir}
    \varphi_{2D}(\rho)= \dot{\iota} \bigl\{J_1(q\rho)-\frac{\dot{\iota}}{4}f_{2D}(\textit{q})H_1(q\rho)\bigl\},
\end{equation}
where $J_1(q\rho)$ and $H_1(q\rho)$ are the Bessel and Hankel functions respectively.
Here, $q$ is the relative wave vector in two dimensions. 
The two dimension scattering amplitude $f_{2D}(q)$ is related to scattering phase shift $\delta(q)$ as $f_{2D} (q)=-4/ (\cot \delta(q)-\dot{\iota} )$.
For low collision energy $E=\hbar^{2} q^{2}/m$, the effective range expansion can be written as $q^{2} \cot \delta (q)=-1/A_p - B_p q^2$~\cite{zhang_signature_2017, zhang_strongly_2017}.
Then the scattering amplitude of $p$-wave is given by \cite{peng_manipulation_2014,zhang_signature_2017},
\begin{equation}\label{amp}
    f_{2D}(q)= \frac{4q^2}{1/A_p + B_pq^2 +\dot{\iota} q^2}.
\end{equation}
Here, $A_p=\frac{3\sqrt{2\pi}l_0^2}{4}\left( \frac{l_0^3}{V_p}+\frac{k_e l_0}{2}-0.065553 \right)^{-1}$ is known as scattering area, which is a controllable interaction parameter. The scattering area depends on three-dimensional scattering volume $V_p$, effective range $k_e$, and harmonic oscillator length $l_0$.
The scattering volume $V_{p}$ depends on external magnetic field and parameterized as $V_{p}=V_{\rm bg} (1+\Delta B /(B-B_{0}))$. 
Here, $V_{\rm bg}$ and $\Delta B$ are the background scattering volume and resonance width in the magnetic field, respectively~\cite{idziaszek_analytical_2009}. Here, $B_0$ is the resonance position in three dimensions.
The positive dimensionless effective range for quasi-two dimension is $B_p= \frac{4}{3\sqrt{2\pi}}{\left(l_0 k_e-0.14641\right)-\frac{2}{\pi} {\ln{\left(l_0 q\right)}}}$. 

The $p$-wave $S$-matrix element is given by $S(q)= \exp$$(2 \dot{\iota} \delta(q))$. 
The $S$-matrix can be extracted from the scattering amplitude as $f_{2D}=2\dot{\iota}[S(q)-1]$, which results the elastic cross-section 
\begin{equation}\label{sige}
    \sigma = {\frac{16 q^3}{\left(1/A_p + B_p q^2\right)^2 +q^4}}.
\end{equation}
The elastic rate coefficient $Q=v \sigma$ becomes
\begin{equation}\label{k2}
    Q= \frac{32\hbar}{m}{\frac{q^4}{\left(1/A_p + B_p q^2\right)^2 +q^4}},
\end{equation}
where  $v=2 \hbar q / m$ is the relative velocity.  

The two-dimensional confinement-influenced resonance occurs at $1/A_{p}=0$ when the scattering cross-section (scattering amplitude) hits maximum. 
This condition occurs at $B^{\prime}=B_{0}-\frac{V_{\rm bg} \Delta B}{l_{0}^{2}} (k_e/2 +0.065553/l_{0})$. The shift in resonance $B^{\prime} > B_{0}$ increases with the increase of confinement (axial frequency) and has been observed in the recent experiment~\cite{waseem_creation_2016}.
Thermally averaged elastic collision rate is given by $\gamma_{el} = \langle n \rangle \langle Q \rangle$, where $\langle n \rangle =\left(\frac{mk_B}{8\pi \hbar^2}\right){\frac{T_F^2}{T}}$ is the two-dimensional mean density for harmonically trapped gas. 
Here, the Fermi temperature $T_{F}$ is related to the number of atoms $N$ and mean trapping frequency $\omega=\sqrt{\omega_x \omega_y}$ of the trap as $k_B T_{F} = (2N)^{1/2} \hbar \omega$.
Whereas, $\langle Q \rangle$ is the averaged elastic collision rate coefficient over thermal Boltzmann distribution
\begin{eqnarray}
\langle Q \rangle &=&\frac{2}{q^2_T} \int^\infty _0 q Q e^{-q^2/q^2_T} dq \nonumber\\
&= & \frac{64\hbar}{mq^2_T}\int^\infty_0  \frac{{d}\left(q^2/2\right) q^4 e^{-q^2/q^2_T}}{{\left(1/A_p+B_pq^2\right)^2}+ q^4}.
\label{ave}
\end{eqnarray}

\begin{figure}[t]
\begin{tabular}{@{}cccc@{}}
\includegraphics[width=3.25in]{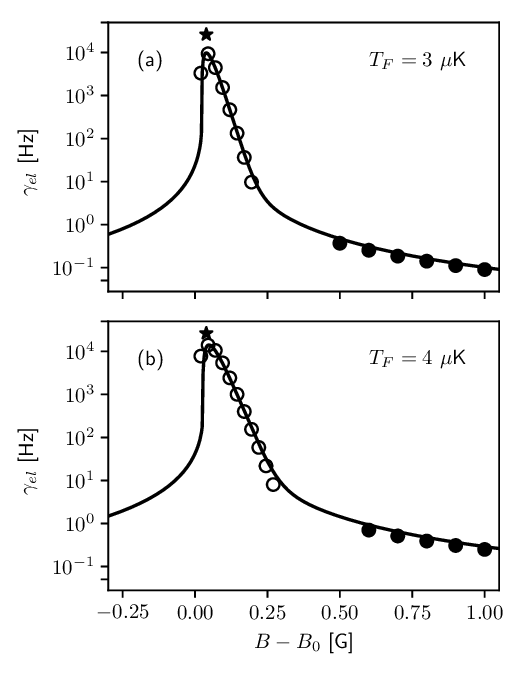}
\end{tabular}
\caption{Numerical results of two dimensional elastic collision rates $\gamma_{el}$ versus magnetic field detuning $B-B_{0}$ at (a) $T_{F}=3~\mu$K and (b) $T_{F}=4~\mu$K, for spin-polarized $^6$Li atoms in lowest hyperfine ground state calculated from  Eq.~(\ref{ave}). Solid markers show off-resonant results from Eq~(\ref{far}) while open circles mark the near-resonant values according to Eq.~(\ref{peak}). These approximate results are in agreement with full numerical results.
The $\star$ points indicate the peak values under the condition of $e^{-q_{r}^{2} / q_{T}^{2}} \approx 1$ in  Eq.~(\ref{peak}).} 
\label{fig:mag2d}
\end{figure}

We, numerically, calculated elastic collision rate $\gamma_{el}$ using Eq.~(\ref{ave}) for $^6$Li atoms at two different temperatures assuming the condition of temperature around Fermi temperature $T=T_{F}$. 
In Fig.~\ref{fig:mag2d}(a) and (b), solid curves show elastic collision rates at $T_{F}=3$~$\mu$K (corresponding density $\langle n \rangle = 1.5 \times 10^{12}/m^{2}$, $\omega_0 = 2 \pi \times 168$~kHz ) and at $T_{F}=4$~$\mu$K (corresponding density $\langle n \rangle = 2.0 \times 10^{12}/m^{2}$, $\omega_0 = 2 \pi \times 194$~kHz). 
The parameters chosen for $^6$Li are $\Delta B=40$G, $V_{\rm bg}=(-41 a_{0})^{3}$ and $k_{e}=0.058/a_{0}$~\cite{nakasuji_experimental_2013, lysebo_ab_2009,zhang_scattering_2010}, here $a_0$ is the Bohr radius.
%%%%%%%%%%%%%%%%%%%%%%%%%%%%%%%%%%%%%%%%%%%%%%%%%%%%%%%%%%%%%%%%%%%%%%%%%%%%%%%%%%%%%%%%%%%%%%%%%%%%%%
%%%%%%%%%%%%%%%%%%%%%%%%%%%%%%%%%%%%%%%%%%%%%%%%%%%%%%%%%%%%%%%%%%%%%%%%%%%%%%%%%%%%%%%%%%%%%%%%%%%%%%%%%%%%

Now we consider the special case of the near resonance regime on the negative side of scattering area $A_{p}<0$. In this case, the two-body system has a resonance pole above the threshold representing the bound state $E_{b}= \hbar^{2} q_{r}^{2}/(2m)$. 
Here, resonant momenta is $q_{r}=1/ \sqrt{B_{p}|A_{p}|}$, which denotes the position of maximum scattering amplitude on the real $q$ and varies with external magnetic field $B$.
In the near resonance regime, the most significant contribution arises from $q_{r}$ under the condition of $q_{T}/q_{r}>1$ or ($A_{p} q_{r}^{2} \ll 1$).
Here, $q_{T}=\sqrt{m k_B T/ \hbar^2}$ is being the thermal momentum.
Near the resonance on its negative side, the largest contribution to the thermally average integral comes from the momenta in a narrow vicinity of $q_r=1/\sqrt{B_p|A_p|}$. This allows us to put $q \equiv q_r$ everywhere except for the first parenthesis in the denominator of Eq.~\ref{ave}, which results
\begin{equation}\label{A3}
\langle Q \rangle^{n}=\frac{32\hbar q^4_r e^{-q^2_r/q^2_T}}{m B_p q^2_T}\int^\infty_{1/{|A_p|}}{\frac{d\left(1/A_p+B_pq^2\right)}{{\left(1/A_p+B_pq^2\right)}^2+q_r^4}}.
\end{equation}
The estimate for the width of the momentum interval $q_r- \delta q \leq q \leq q_r + \delta q$ provides the dominant contribution to the integral. For $q = q_r + \delta q$ we have 
\begin{equation*}
    1/|A_p| + B_p q^2 \sim q^2 \implies B_p \delta q q_r \sim q_r^2.
\end{equation*}
So by dividing the above equation with $q_r^2$ and taking into account that $\delta q \ll q_r$ (for the narrow interval) we get, $\delta q/q_r \sim (|A_p| q_r^2)\ll 1$.
Defining $x = \left(1/A_p+B_pq^2\right)/\alpha$ and $\alpha=q^2_r$, we obtain
\begin{equation}\label{A4}
 \langle Q \rangle^{n}=\frac{32\hbar q^4_r}{mB_pq^2_T}e^{-q^2_r/q^2_T}\frac{1}{\alpha}\int^\infty_{1/\left(|A_p|\alpha \right)}{\frac{dx}{x^2+1}}.
\end{equation}
Using the property of integrals
\begin{equation*}
\int^\infty_{1/\left(|A_p| \alpha \right)}\frac{dx}{x^2+1}  =  \frac{\pi}{2}+\tan^{-1} (1/(|A_p|\alpha))  \approx  \pi,
\end{equation*}
where we assume that near resonance $A_{p} \alpha \ll 1$ such that one can write $\tan^{-1}((1/(|A_p|\alpha))) \approx \pi/2$. 
This results in the approximate expression of the elastic collision coefficient in the near resonance regime, which is given by
\begin{equation}
\langle Q \rangle^{n}=\frac{32\pi\hbar}{mB_p}{\left(\frac{q_{r}}{q_T}\right)^2} e^{-q_{r}^{2} / q_{T}^{2}}.
\label{peak}
\end{equation}
Since, $\langle Q \rangle^{n} \propto q_{r}^2 \propto l_0/V_{p}$. Therefore, tight harmonic confinement has a nearly linear influence on $\langle Q\rangle^{n}$.
The open circles in Fig.~\ref{fig:mag2d} show the estimated values of elastic collision rate $\gamma_{el}= \langle n \rangle \times \langle Q\rangle^{n}$, which agrees well with direct numerical simulations in the close vicinity of the resonance where elastic collision rates enhanced by four orders of magnitude with exponential scaling of Eq.~\ref{peak}.
If $q_{T}/q_{r}$ is much larger than unity, which can be achieved with high temperature and at very small magnetic field detuning.
As a result, one can expand the exponent in Eq.~(\ref{peak}) in Taylor series $e^{-q_{r}^{2} / q_{T}^{2}} \approx 1$, which gives the peak value of elastic collision rate indicated by $\star$ points in Fig.~\ref{fig:mag2d}. This peak value is slightly overestimated. Combining with $\langle n \rangle= q_{T}^{2}/(8 \pi)$, it turns out that the peak becomes nearly independent of temperature, this condition can be well satisfied at higher temperatures.

In the experiment, elastic collision rates can be extracted from thermalization rate $\Gamma_{th}=\gamma_{el}/ \alpha$.
Here, $\alpha \approx 4.1$ is an average number of $p$-wave elastic collisions needed for time evolution for the temperature difference between two radial directions towards thermalization in two dimensional geometry~\cite{zhu_evaporative_2013}.
Thermalization rates can be directly extracted using a cross-dimensional relaxation method similar to three-dimensional $p$-wave case~\cite{top_spin-polarized_2021,nakasuji_experimental_2013, monroe_measurement_1993}.
In order to benchmark the approximate expression (\ref{peak}), we have also derived a similar expression for the three-dimensional case in Appendix A. 
Then we compare it to experimental data of thermalization rate from Nakasuji \textit{et al.},~\cite{nakasuji_experimental_2013} and found a good agreement (for detail see Appendix A).

Next, we consider the weakly interacting regime which is sufficiently far from the resonance ($A_{p} \rightarrow 0$). In this regime, ratio $q_{T}/q_{r}$ is very low approximately 0.3 in the experiments~\cite{yoshida_scaling_2018, waseem_quantitative_2019}.
In this zero-range limit when the effective range is negligible, the dominant contribution arises from $1/A_p$ in the denominator of Eq.~(\ref{k2}). After averaging over the thermal Boltzmann distribution~(Eq.~\ref{ave}), estimated value of the elastic rate coefficient becomes
\begin{equation}
\langle Q \rangle^f= {\frac{64 m A_{p}^2}{\hbar^3}}{\left(k_B T\right)^2},
\label{far}
\end{equation}
which shows quadratic temperature dependence. 
The filled circles in Fig.~\ref{fig:mag2d} show the estimated far resonance value of elastic collision rate $\gamma_{el}= \langle n \rangle \times \langle Q\rangle^{f} $, which agrees with numerical results with very slight underestimation.
Eqs.~\ref{peak} and~\ref{far} indicate that  elastic collision rates exhibit different temperature dependencies in near and far resonance cases. 
%===============================================================
\begin{figure}[t]
\begin{tabular}{@{}cccc@{}}
\includegraphics[width=3.25 in]{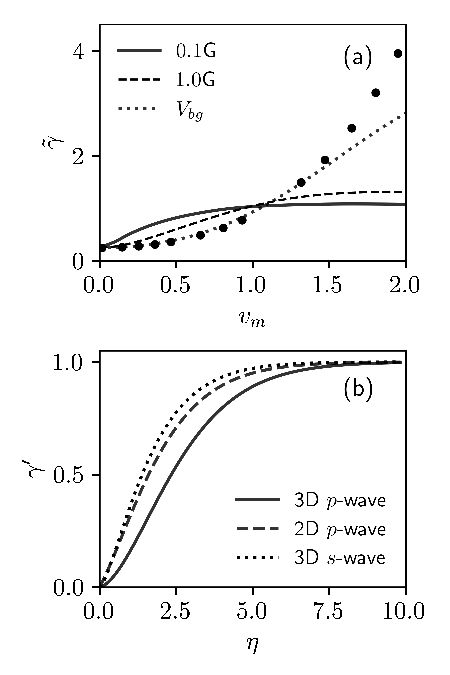}
\end{tabular}
\caption{(a) Numerical results of normalized $p$-wave evaporation rates $\Tilde{\gamma}$ versus mean speed $v_{m}$. 
Solid, dashed, and dotted curves show the result at $B-B_{0}=0.1$G, $B-B_{0}=1.0$G, and $B-B_{0}=V_{bg}$, respectively.
Filled circles show the results of Eq.~\ref{coll2}, which agrees with numerical results at low velocity.
(b) The fraction of collisions up to a kinetic energy of $\eta k_BT$, where $\eta$ is the truncation parameter.}
\label{fig:Evap}
\end{figure}
%===============================================================

%=========================================================================
\section{Energy transfer from high to low velocity}
%==================================
In the previous section, we have observed that elastic collision rates are significantly enhanced up to fourth order of magnitude on approaching to the zero magnetic field detuning and well described by Eq.~\ref{peak}.
However, in the case of spin-polarized fermions, at the same time, three-body losses are also significantly enhanced with the same scaling of $e^{-q_{r}^{2} / q_{T}^{2}}$ towards zero magnetic fields detuning as observed in a recent experiment in two dimension~\cite{waseem_quantitative_2019}, which significantly reduce the lifetime of gas.
On the other hand, $p$-wave interactions are very weak at low temperatures and at large magnetic field detuning (i.e., near background $p$-wave collision in the ultracold regime).
Therefore, in this section, we theoretically explore to what extent the $p$-wave elastic collisions influence the evaporation process by increasing the density or temperature of the gas until three-body recombination becomes too strong.
In evaporative cooling, after removing high-velocity atoms from the tail of the Maxwell-Boltzmann distribution, atoms transfer energy from high to low velocity in order to recover the Maxwell-Boltzmann distribution.
Estimating the required number of $p$-wave collisions to recover the Maxwell-Boltzmann distribution is an important benchmark.
The collision rate with speed $v$ involving an atom is
\begin{equation}
\label{coll}
\gamma_{|v_{1}=v|} = \langle n\rangle \int \sigma |v-v_2| f_{v_2} d^2 v_2,
\end{equation}
where $\sigma$ is the collision cross-section, $f_{v_2}$ is the two dimensional Maxwell-Boltzmann velocity distribution, and $|v-v_2| = \left( v^2+v_2^2-2vv_2\cos\theta \right )^{1/2}$.
Introducing, $\tilde{u}^2=\frac{1}{2} \beta m v^2$ and $u^2=\frac{1}{2} \beta m v_{2}^2$ with $\beta = 1/ (k_B T)$, Eq.~\ref{coll} becomes
\begin{equation}
\label{coll2}
\gamma_{|v_{1}=v|}= \frac{\langle n \rangle}{\pi} \left(\frac{2}{\beta m} \right)^{1/2} \int \sigma(\Tilde{u}-u) |\Tilde{u}-u| e^{-u^2} u du d\theta.
\end{equation}
Here, elastic cross section $\sigma$ is $\sigma (q) \equiv \sigma(|\Tilde{u}-u|)$. 
The wave vector $q$ in terms of velocity is $q=\sqrt{m/2\beta \hbar^2} |\Tilde{u}-u|$. 
We define relevant dimension less quantity $\tilde{\gamma}=\gamma_{|v_{1}=v|}/\gamma_{el}$ to estimate the elastic collision rate ($p$-wave evaporation rate) for an atom with velocity $v$. 
In Fig.~\ref{fig:Evap}(a), we show two-dimensional $p$-wave normalized evaporation rates as a function of mean velocity $v_{m}=\sqrt{\pi k_BT/2m}$ at three different magnetic field detunings using direct numerical results from Eq.~\ref{coll2} and \ref{ave}.   
Solid, dashed, and dotted curves show the result at $B-B_{0}=0.1$G, $B-B_{0}=1.0$G, and $B-B_{0}=V_{bg}$, respectively.

In the region of low temperature limit when mean velocity is close to zero, the elastic scattering cross-section from Eq.~\ref{sige} becomes $\sigma(\Tilde{u}-u)=16 A_{p}^2 (\sqrt{m/2\beta \hbar^2} |\Tilde{u}-u|)^3$. In this limit, dimensionless quantity $\tilde{\gamma}$ reduces in the following form 
\begin{equation}\label{collnorm}
\tilde{\gamma} = \frac{\int_0^\infty\int_0^{2\pi}\left(\tilde{u}^2+u^2-2\tilde{u}u \cos\theta \right)^2 d\theta u e^{-u^2}}{8\pi}  du.
\end{equation}
Filled circles in Fig.~\ref{fig:Evap}(a) shows the results of above equation, which agree well in low velocity limit at $V_{p}\approx V_{\rm bg}$. 
At higher velocities, the zero-range limit breaks down due to increase of density or temperature. 
Therefore, the results of Eq.~\ref{collnorm} starts to deviate from numerical results.   
For the same averaged collision rates, transfer of energy to temperature range near $T=0$ (or velocity group $v=0$) is $\sqrt{2}$ times to the $p$-wave collisions in two dimensions compared with the $p$-wave case in three dimensions (as estimated in Appendix A).

The decrease in temperature from an initial temperature depends on the truncation of energy distribution at $\eta k_B T$ followed by thermal relaxation in an infinitely deep potential~\cite{davis_analytical_1995}. Here, $\eta$ is the truncation parameter.
We calculate fraction of collision rates by thermally averaging over all possible speeds, given by
\begin{equation}\label{frac}
\gamma^{\prime} =\int_0^{v^{\prime}} \gamma_{|v_{1}=v|} f_{v} dv.
\end{equation}
Here, upper limit is $v^{\prime}=\sqrt{\eta  (2 k_B T/ m)}$ and $f_{v}$ is the Maxwell-Boltzmann speed distribution. 
It is parameterized for three dimension as $f_{v} = 4\pi (\beta m/2\pi)^{3/2} v^2 e^{- \beta m v^2/2}$ while for two dimension it is $f_{v} = (\beta m) v e^{- \beta m v^2/2}$.
In Fig.~\ref{fig:Evap} (b) dashed curve shows the fraction of $p$-wave collision rates as a function of $\eta$ for the two-dimensional case.
The solid curve represents the three-dimensional $p$-wave collision rates fraction while the black dotted curve represents the three-dimensional $s$-wave collision rates fraction~\cite{top_spin-polarized_2021}.
Typically, for $\eta < 8$, truncation effects become important~\cite{luiten_kinetic_1996}.
For small $\eta$ less than 4, the fraction of atoms that can escape over the threshold of potential is large in two dimensions compared to three dimensions $p$-wave case. In other words, its approaches more closely to the $s$-wave case.
However, the temperature reduction per escape atom is usually small~\cite{davis_analytical_1995}. Therefore, the interplay between the truncation parameter and the dynamic of evaporation suggests that $\eta$ should be kept large and constant~\cite{wilkowski_runaway_2010}. 
In the range of $6 \geq \eta \leq 8$, two dimensions $p$-wave collision rates fraction gets much closer to three dimensions $s$-wave collision rates compared to three dimensions $p$-wave case. In this range, residual evaporation balance the heat generated by inelastic collision losses~\cite{rem_lifetime_2013}.
%%%%%%%%%%%%%%%%%%%%%%%%%%%%%%%%%%%%%%%%%%%%%%%%%%%%%%%%%%%%%%%%%%%%%%%%%%%%%%%%%%%%%%%%%%%%%%%%%%%%%%
\begin{figure}[t]
\begin{tabular}{@{}cccc@{}}
\includegraphics[width=3.25 in]{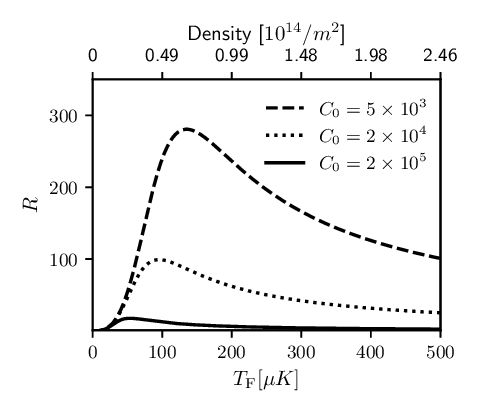}
\end{tabular}
\caption{Ratio of two-body elastic to three-body inelastic collisions for two-dimensional harmonically trapped $^6$Li atoms near zero field or far away from the $p$-wave Feshbach resonance with $T=T_{F}$ assuming vacuum life time of 60 seconds. The favorable region is the lower value of dimensionless three-body loss constant $C_0$ and at some intermediate densities.}
\label{fig: ratio2d}
\end{figure}
%%%%%%%%%%%%%%%%%%%%%%%%%%%%%%%%%%%%%%%%%%%%%%%%%%%%%%%%%%%%%%%%%%%%%%%%%%%%%%%%%%%%%%%%%%%%%%%%%%%%%%%%%%%%

%===============================================================================================
\section{Collisional stability}
%===============================================================================================
Next, we analyze the collision stability in the weakly interacting regime. 
There is no inelastic two-body collision for identical fermions in the lowest hyperfine ground state due to angular momentum conservation, especially for the $^6$Li atom~\cite{zhang_p-wave_2004,waseem_two-body_2017}.
In spin-polarized fermions, in the lowest hyperfine ground state, dominant inelastic losses are only due to a three-body inelastic collision.
Therefore, we only focus on the ratio of two-body elastic (good) to three-body inelastic (bad) collisions in spin-polarized fermions confined in two dimensions.
Good-to-bad collisions ratio $R$ is generally defined as 
\begin{equation}
R=\frac{\gamma_{el}}{\gamma_{in}+(1/ \tau )}= \frac{\langle n \rangle \times \langle Q \rangle }{\langle n^2 \rangle \times Q_{3}+ (1/ \tau) }
\label{r}
\end{equation}
where $\tau$ is the vacuum-limited lifetime of atoms in a trap and $Q_3$ is the three-body inelastic loss coefficient.
Here, $\langle n^{2} \rangle = \frac{1}{48} \left(\frac{mk_B}{\pi\hbar^2}\right)^2 \frac{T_F^4}{T^2}$ is the mean squared density for two dimensional harmonically trapped gas.

Considering the extreme case where the scattering area is quite small almost closer to the background scattering. We assume that density exceeds from $1/ \lambda^{2}$ ($n \approx 10^{12}/m^{2}$).
Here, $\lambda$ is the wavelength of resonant light. 
In this zero-range limit, the effective range $B_{p}$ can be assumed negligible~\cite{yoshida_scaling_2018,suno_recombination_2003,top_spin-polarized_2021}.
Elastic cross-section is scaled as $\sigma \propto A_{p}^{2} q^{3}$ (as evident from Eq.~\ref{far}). 
At temperature $T=T_{F}$, elastic collision rate $\gamma_{el}$ is proportional to $A_{p}^{2} q^{6}$.
In the zero-range approximation, for sufficiently small scattering area $A_{p}$, three-body inelastic loss coefficient parameterized through scaling relation \cite{waseem_quantitative_2019}.
\begin{equation}
Q_{3}= C_{0}  \frac{\hbar}{m} q_{T}^{4} A_{p}^{3},
\label{q3}
\end{equation}
with dimensionless non-universal scaling constant $C_0$.
Hence, three body loss rate $\gamma_{in}= \langle n^2 \rangle \times Q_{3}$ becomes proportional to $A_{p}^{3} q^{8}$.
 This implies that the ratio of good-to-bad collision scales as $1/ (C_0 A_{p} q^{2})$. Near zero field when the interaction is closer to the background scattering, the collision ratio mainly depends on achievable densities ($q$ or $T_{F}$) and constant $C_0$.
Therefore, a favorable regime is at low densities with weak $p$-wave interactions.

Scaling of Eq.~\ref{q3} has been measured in the experiment with $C_0=2 \times 10^4$ around 0.5G detuning~\cite{waseem_quantitative_2019}.
In the case of spin-polarized fermions in three dimensions, the equivalent scaling law is $K_{3}=C (\hbar/m) k^{4} V_{p}^{8/3}$~\cite{suno_recombination_2003} and has been observed in recent experiment~\cite{yoshida_scaling_2018, top_spin-polarized_2021}. Here, $k$ is the thermal wave vector and $C$ is the equivalent dimensionless constant in three dimensions.  The value of $C=2 \times 10^6$ was reported around 0.5~G detuning in Ref.~\cite{yoshida_scaling_2018}.
Ref.~\cite{top_spin-polarized_2021}  reported two order smaller value $C=3 \times 10^4$ around 1~G detuning. 
These different $C$ values indicate the lack of universal character in the recombination of three ultracold fermions~\cite{suno_three-body_2003, suno_recombination_2003}, which means $C$ depends on the detail of interatomic potential unlike Bose gases~\cite{nielsen_low-energy_1999}. 
Therefore, the value of $C_0$ or $C$ might vary with scattering strength, and the value of $C_0$ is of crucial importance for the quantitative evaluation of the three-body losses. But its value is not well defined. 
Therefore, considering the non-universal character of $C_{0}$, we calculate the ratio of good-to-bad collision near background $p$-wave collision as a function of temperature at three different values of $C_0$. The results are shown in Fig.~\ref{fig: ratio2d} where dashed curve is for $C_0=5\times 10^3$, dotted curve is for $C_0=2 \times 10^4$, and solid curve is for $C_0=2\times 10^5$.
Here, we assumed the vacuum limited lifetime $\tau \approx 60$~seconds.
The good-to-bad collision ratio can lead to a favorable regime at somewhere intermediate densities for the lower value of $C_0$.
There is a favorable range around $T_{F} \approx 100-200$~$\mu$K where the ratio of good-to-bad collision reaches maximum 100 to 300. In this favorable range, corresponding densities range is from $\langle n \rangle \approx 0.5\times 10^{14}/ m^{2}$ to $\langle n \rangle \approx 1.0 \times 10^{14}/ m^{2}$, and $C_0$ values range is from $C_0 \approx 2 \times 10^4$ to $C_0 \approx 5 \times 10^3$.

Next, it is natural to focus on a close resonance regime where the scattering area is maximum and the condition of $q_{T}/q_{r} \geq 1$ is well satisfied.
In this regime, the three-body loss coefficient $Q_{3}$ shows the unitary behavior. In the unitary regime, the three-body loss coefficient depends only on temperature and no dependence on scattering area.
Considering $q_{T}$ as the only relevant length scale in the unitary limit, expected scaling is $Q_{3} \propto (\hbar / m) q_{T}^{-2}$ from the dimensional analysis. After taking the thermal average over Boltzmann distribution, we get the maximum upper limit of three-body loss constant\cite{dincao_ultracold_2015}
\begin{equation}
Q_{3}^{p}= \zeta \frac{3 \pi \hbar^{3}}{m^{2} (k_B T)}=\zeta \frac{3 \pi \hbar}{m q_{T}^2}.
\label{unit}
\end{equation}
Here, we introduce the specie dependant non-universal dimensionless constant $\zeta \leq 1$ similar to three dimension case~\cite{waseem_unitarity-limited_2018, rem_lifetime_2013, fletcher_stability_2013}. 
From Eq.~\ref{peak} The peak value of the elastic collision rate coefficient is proportional to $1/(A_{p} q_{T}^2)$.
Assuming Fermi gas at temperature $T=T_{F}$, $\langle n \rangle / \langle n^2 \rangle= 6 \pi / q_{T}^2$. This implies that the ratio of elastic to inelastic collision rate (in units of Hz) is nearly scaled to $1/(\zeta A_{p} q_{T}^2)$. 
Since $A_{p}$ is quite large at resonance, it results in a low value of good-to-bad collision ratio. However, tailoring achievable density (or $T_{F}$) to lower value improves the good-to-bad collision ratio, as it is possible in two dimensions.
This suggests that thermalization measurement similar to the three-dimensional case can be performed with better resolution in experiments and also some new interesting few-body physics.

%================================================================
\section{Conclusion}
%================================================================
In summary, we studied the elastic collision rates in two regimes of $p$-wave interactions for two-dimensional spin-polarized fermions in the lowest hyperfine ground state of $^6$Li atoms. 
In the non-zero-range limit where the effective range term is dominant, elastic collision rates are enhanced exponentially towards the resonance with scaling of $e^{-q_{r}^{2} / q_{T}^{2}}$ and also in agreement with direct numerical results. 
The derived analogous expression for the case of three dimensions successfully explains the experimental results of thermalization rates measured by Nakasuji et al., ~\cite{nakasuji_experimental_2013}. 
In the zero-range limit, when the effective range is negligible and the scattering area $A_p$ is near the background $p$-wave collision in the ultracold regime, the elastic collision rate coefficient is proportional to temperature $T^2$ and scattering area $A_{p}^2$. 
In this background limit, the transfer of energy from high to low velocities through $p$-wave collision is almost $\sqrt{2}$ times than the three-dimensional case, as long as thermal averaging condition holds. Such condition can be control in experiment using tight confinment such that collision of fermions can not populate the excited transversal states~\cite{waseem_quantitative_2019, kurlov_two-body_2017}. 
We would like to mentioned that our analysis is based on single channel calculation which provided useful insight of p-wave collision dynamics in 2D. However, further detail insight requires multichannel treatment in order to assess the cross-over from 3D to 2D, Pauli Blocking effects and its impact on evaporative cooling.

%\setcounter{equation}{0}
%\renewcommand{\theequation}{A\arabic{equation}}
%==========================================================================================================

%\section{Appendixes}

%\begin{verbatim}
\appendix
\section{Elastic collision rates in three dimensions}
%################################################
%----------------------------------------------------------
\begin{figure}[t]
\includegraphics[width=3.2 in]{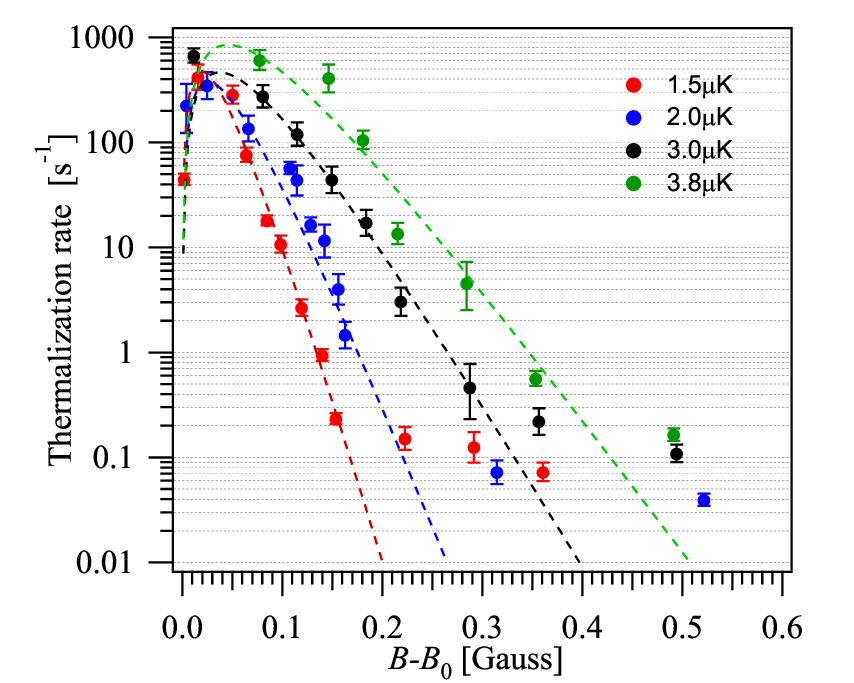} 
\caption{Thermalization rates close to $p$-wave Feshbach resonance at $B-B_0=159.17(5)G$ from publication of Nakasuji \textit{et al.},~\cite{nakasuji_experimental_2013} for four different sets of temperatures.
The dashed curves show the result obtained from Eq.~\ref{therm3} with all fixed scattering parameters, which successfully reproduces the experimental results in a narrow range closer to the resonance.}
\label{fig:naka} 
\end{figure}
%----------------------------------------------------------
%%%%%%%%%%%%%%%%%%%%%%%%%%%%%%%%%%%%%%%%%%%%%%%%%%%%%%%%%%%%
%--------------------------------------------------------
%\end{verbatim}
The scattering amplitude for $p$-wave interaction between two fermions with relative wave vector $k$ in three dimension is given by
\begin{equation}\label{amp3}
f(k)=\frac{-k^2}{1/V_{p}+k_e k^2+i k^3}.
\end{equation}
Here, $V_{p}$ is scattering volume and $k_e > 0$ is the effective range. 
The $p$-wave $S$-matrix element is given by $S(k)= \exp$$(2 i \delta(k))$. The elastic rate constant is $K=v \sigma(k)$, 
where $\sigma(k)=3 \pi \left|1-S(k)\right|^{2} /k^{2} $ is the $p$-wave elastic scattering cross section, and $v=2 \hbar k / m$ is the relative velocity. As a result, the elastic rate coefficient becomes
\begin{equation}
K= {\frac{24\pi \hbar}{m}}{\frac{k^5}{(1/V_{p}+k_e k^2)^2 + k^6}}.
\label{K3}
\end{equation}
Very close to the resonance when $V_{p} \rightarrow \infty$, the largest contribution comes from the momenta of resonant bound state $k_r= 1  /\sqrt{k_e|V_{p}|}$. 
In the close vicinity of resonant regime, $k_{T} \gg k_r$, where $k_{T}=\sqrt{3m k_BT/2\hbar^2}$ is the thermal momentum. As a result, only a small fraction of relative momenta contributes to the collision process.
Following the procedure similar to the two-dimensional case in the main text, we find the expression for the elastic collision rate coefficient 
\begin{equation}
\langle K\rangle^n=\frac{96{\pi}^{3/2}\hbar}{m k_e}{\left(\frac{k_r}{k_{T}}\right)^3} e^{-\left(k^2_{r}/k^2_{T} \right)}.
\label{k3peak}
\end{equation}
Thermalization rates can be obtained from the above equation as 
\begin{equation}
\Gamma_{th}= \langle n \rangle \times \langle K\rangle^{n}/ \alpha.
\label{therm3}
\end{equation}

The mean density for a three-dimensional harmonically trapped Fermi gas at temperature $T$ in the Boltzmann regime is given by $\left\langle n \right\rangle={\frac{1}{48}}{\left(\frac{mk_B}{{\hbar^2\pi}}\right)^{3/2}} {\frac{T_F^3}{T^{3/2}}}$~\cite{top_spin-polarized_2021}.
The dashed curves in Fig.~\ref{fig:naka} show the fitted thermalization rates $\Gamma$ in comparison to the experimental data from Ref.~\cite{nakasuji_experimental_2013} for four different sets of temperatures. During the fitting we kept all scattering parameters fixed and kept the mean density as the only free parameter. The expression~\ref{therm3} successfully reproduces the experimental results in the narrow range where interaction is sufficiently strong.
The mean density obtained from fitting differs from the measured density of approximately $50 \%$ due to uncertainty in the estimation of atom numbers in the trap and as well as trap conditions such as trapping frequencies.

At sufficiently far away from the resonance, interaction is weak ($V_{p} \rightarrow 0$) and $k_{T} \ll k_r$.
In this regime, elastic collision rates can be approximated as~\cite{top_spin-polarized_2021}
\begin{equation}
\Gamma^f= \left\langle n \right\rangle \times \frac{288\sqrt{\pi}V_{p}^2 m^{3/2}}{{\hbar}^4} { \left(k_BT \right)}^{\frac{5}{2}}. 
\label{k3far}
\end{equation}
The ratio of elastic scattering rate for an atom with velocity $v$ compared to average scattering rate $\Gamma^{f}$~\cite{top_spin-polarized_2021}:
\begin{equation}\label{collnorm3d}
\tilde{\Gamma} =\frac{\int_0^\infty\int_0^{\pi} \left(\tilde{u}^2+u^2-2\tilde{u}u \cos \theta\right)^{5/2} \sin\theta d\theta u^2 e^{-u^2} du}{24 \sqrt{3}} 
\end{equation}
At zero velocity one can substitute $v=0$ ($\tilde{u}=0$) in Eq.~\ref{collnorm3d} and in Eq.~\ref{collnorm}, which results the ratio $\tilde{\gamma}/\tilde{\Gamma}=\sqrt{2}$.

%===========================================================================================================

\section*{ACKNOWLEDGEMENT}
We acknowledge the fruitful discussions with Yair Margalit.

%\clearpage
%\begin{thebibliography}{90}
\bibliographystyle{apsrev4-2}
\bibliography{Ref.bib}

%\end{thebibliography}

\end{document}